# Tailoring single-cycle electromagnetic pulses in the 2–9 THz frequency range using DAST/SiO$_2$ multilayer structures pumped at Ti:sapphire wavelength

Andrei G. Stepanov,[1,2*] Andrii Rogov,[2] Luigi Bonacina,[2] Jean-Pierre Wolf,[2] and Christoph P. Hauri[1,3]

*[1]Paul Scherrer Institute, 5232 Villigen, Switzerland*
*[2]GAP-Biophotonics, Université de Genève, Chemin de Pinchat 22, CH - 1211 Genève 4, Switzerland*
*[3]Ecole Polytechnique Federale de Lausanne, 1015 Lausanne, Switzerland*
*[*]andrey.stepanov@psi.ch*

**Abstract:** We present a numerical parametric study of single-cycle electromagnetic pulse generation in a DAST/SiO$_2$ multilayer structure via collinear optical rectification of 800 nm femtosecond laser pulses. It is shown that modifications of the thicknesses of the DAST and SiO$_2$ layers allow tuning of the average frequency of the generated THz pulses in the frequency range from 3 to 6 THz. The laser-to-THz energy conversion efficiency in the proposed structures is compared with that in a bulk DAST crystal and a quasi-phase-matching periodically poled DAST crystal and shows significant enhancement.

## References and Links

1. M. Schirmer, M. Fujio, M. Minami, J. Miura, T. Araki, and T. Yasui, "Biomedical applications of a real-time terahertz color scanner," Biomed. Opt. Express **1**(2), 354–366 (2010).
2. H. Y. Hwang, S. Fleischer, N. C. Brandt, B. G. Perkins Jr., M. Liu, K. Fan, A. Sternbach, X. Zhang, R. D. Averitt, and K. A. Nelson, "A review of non-linear terahertz spectroscopy with ultrashort tabletop-laser pulses," J. Mod. Opt. DOI: 10.1080/09500340.2014.918200.
3. C. Vicario, C. Ruchert, F. Ardana-Lamas, P. M. Derlet, B. Tudu, J. Luning, and C. P. Hauri, "Off-resonant magnetization dynamics phase-locked to an intense phase-stable terahertz transient," Nature Photon. **7**(9), 720–723 (2013).
4. N. Stojanovic and M. Drescher, "Accelerator- and laser-based sources of high-field terahertz pulses," J. Phys. B: At. Mol. Opt. Phys. **46**(19), 192001–192013 (2013).
5. E. Roman, J. R. Yates, M. Veithen, D. Vanderbilt, and I. Souza, "Ab initio study of the nonlinear optics of III-V semiconductors in the terahertz regime," Phys. Rev. B **74**(24), 245204 (2006).
6. A. G. Stepanov, S. Henin, Y. Petit, L. Bonacina, J. Kasparian, and J.-P. Wolf, "Mobile source of high-energy single-cycle terahertz pulses," Appl. Phys. B **101**(1-2), 11–14 (2010); A. G. Stepanov, S. Henin, Y. Petit, L. Bonacina, J. Kasparian, and J.-P. Wolf, "Erratum to: Mobile source of high-energy single-cycle terahertz pulses," Appl. Phys. B **115**(2), 293 (2014).
7. J. A. Fülöp and J. Hebling, *Applications of Tilted-Pulse-Front Excitation; in Recent Optical and Photonic Technologies*, edited by K. Y. Kim (NTECH, 2010).
8. J. Hebling, A. G. Stepanov, G. Almasi, B. Bartal, and J. Kuhl, "Tunable THz pulse generation by optical rectification of ultrashort laser pulses with tilted pulse fronts," Appl. Phys. B **78**(5), 593–599 (2004).
9. C. Vicario, B. Monoszlai, Cs. Lombosi, A. Mareczko, A. Courjaud, J. A. Fülöp, C. P. Hauri, "Pump pulse width and temperature effects in lithium niobate for efficient THz generation," Opt. Lett. **38**(24), 5373–5376 (2013).
10. S.-W. Huang, E. Granados, W. R. Huang, K.-H. Hong, L. E. Zapata, and F. X. Kärtner, "High conversion efficiency, high energy terahertz pulses by optical rectification in cryogenically cooled lithium niobate," Opt. Lett. **38**(5), 796–798 (2013).
11. K.Y. Kim, J.H. Glownia, A. J. Taylor, and G. Rodriguez, "High-power broadband terahertz generation via two-color photoionization in gases," IEEE J. Quantum Electron. **48**(6), 797–805 (2012).
12. C. P. Hauri, C. Ruchert, C. Vicario, and F. Ardana, "Strong-field single-cycle THz pulses generated in an organic crystal," Appl. Phys. Lett. **99**(16), 161116 (2011).

13. C. Ruchert, C. Vicario, and C. P. Hauri, "Scaling submillimeter single-cycle transients toward megavolts per centimeter field strength via optical rectification in the organic crystal OH1," Opt. Lett. **37**, 899–901 (2012).

14. C. Ruchert, C. Vicario and C. P. Hauri, "Spatiotemporal focusing dynamics of intense supercontinuum THz pulses," Phys. Rev. Lett. **110**(12), 123902 (2013).

15. C. Vicario, C. Ruchert and C. P. Hauri, "High field broadband THz generation in organic materials," J. Mod. Opt. 1–6 (2013).

16. C. Vicario, B. Monoszlai, and C. P. Hauri, "GV/m single-cycle terahertz fields from a laser-driven large-size partitioned organic crystal." Phys. Rev. Lett. **112**(21), 213901 (2014).

17. B. Monoszlai, C. Vicario, M. Jazbinsek, and C. P. Hauri, "High-energy terahertz pulses from organic crystals: DAST and DSTMS pumped at Ti:sapphire wavelength," Opt. Lett. **38**(23), 5106–5109 (2013).

18. A. G. Stepanov, L. Bonacina, and J.-P. Wolf, "DAST/SiO$_2$ multilayer structure for efficient generation of 6 THz quasi-single-cycle electromagnetic pulses," Opt. Lett. **37**(13), 2439–2441 (2012).

19. J. A. Armstrong, N. Bloembergen, J. Ducuing, and P. S. Pershan, "Interactions between light waves in a nonlinear dielectric," Phys. Rev. **127**(6), 1918–1939 (1962).

20. Y.-S. Lee, T. Meade, V. Perlin, H. Winful, T. B. Norris, and A. Galvanauskas, "Generation of narrow-band terahertz radiation via optical rectification of femtosecond pulses in periodically poled lithium niobate," Appl. Phys. Lett. **76**(18), 2505–2507 (2000).

21. F. Pan, G. Knöpfle, C. Bosshard, S. Follonier, R. Spreiter, M. S. Wong, and P. Günter, "Electro-optic properties of the organic salt 4-N,N-dimethylamino-4-N-methylstilbazolium tosylate," Appl. Phys. Lett. **69**(1), 13–15 (1996).

22. P. D. Cunningham and L. M. Hayden, "Optical properties of DAST in the THz range," Opt. Express **18**(23), 23620–23625 (2010).

23. G. Ghosh, "Dispersion-equation coefficients for the refractive index and birefringence of calcite and quartz crystals," Opt. Commun. **163** (1-3), 95-102 (1999)

24. E. E. Russel and E. E. Bell, "Measurement of the optical constants of crystal quartz in the far infrared with the asymmetric Fourier-transform method," JOSA **57**(3), 341–348 (1967).

25. M. Thakur, J. Xu, A. Bhowmik, and L. Zhou, "Single-pass thin-film electro-optic modulator based on an organic molecular salt," Appl. Phys. Lett. **74**(5), 635-637 (1999).

26. S. Brahadeeswaran, Y. Takahashi, M. Yoshimura, M. Tani, S. Okada, S. Nashima, Y. Mori, M. Hangyo, H. Ito, and T. Sasaki, "Growth of ultrathin and highly efficient organic nonlinear optical crystal 4′-dimethylamino-N-methyl-4-stilbazolium p-chlorobenzenesulfonate for enhanced terahertz efficiency at higher frequencies," Cryst. Growth Des. **13**(2), 415–421 (2013).

## 1. Introduction

High-energy ultrashort THz pulses are of great current interest owing to their applications in biomedical imaging [1], THz nonlinear spectroscopy [2] including ultrafast magnetization dynamics study [3], and characterization of ultrashort electron and X-ray pulses [4]. In spite of the significant progress that has been made in the development of THz techniques during the last 10 years, generation of powerful single-cycle THz pulses in the 3–9 THz range is still a challenging task. Intense THz pulses in this frequency range are required for many scientific purposes, for example, for studying the nonlinear susceptibility of polar semiconductors and dielectrics in the vicinity of the fundamental lattice vibrations [5]. At present, high-energy (up to 175 μJ, delivered by a mobile THz source [6]) near-single-cycle pulses with an average frequency below 1.5 THz can be generated by optical rectification of femtosecond laser pulses with tilted pulse fronts in a lithium niobate crystal [7]. Attempts to use this technique for obtaining THz pulses with higher frequencies result in a significant drop in the conversion efficiency [8], which mainly occurs due to the strong increase of THz absorption in lithium niobate above 2 THz even at cryogenic temperatures [9, 10]. Near-single-cycle THz pulses in the frequency range of 3–9 THz can be generated by two-color photoionization in gases [11] but in this frequency range, the laser-to-THz conversion efficiency is not high enough to provide generation of high-energy THz pulses. Organic crystals have recently been shown to emit efficiently across the 1-10 THz range [12-17]. Pumped with a near infrared laser (1.3-1.5 μm) excellent phase-matching provides high conversion efficiency (1-2%) and large THz pulse energies (<50 μJ).

In practice, a pump scheme based just on a Ti:sapphire amplifier is technically easier to handle than a pump at 1.5 μm as the latter requires an additional nonlinear frequency conversion stage (typically a multi-stage optical parametric amplifier). This leads to a significant increase of the overall complexity and source instabilities. Recently broadband terahertz pulses were demonstrated from DAST and DSTMS crystals pumped directly at Ti:sapphire wavelength [17]. Unfortunately the laser-to-THz energy conversion efficiency is about 2 orders of magnitude lower than for a 1.5 μm pump as a different crystal orientation is required for phase-matching for which the element $\chi_{122}^{OR}$ of the nonlinear susceptibility tensor is employed. The $\chi_{122}^{OR} = 166$ pV/m in the DAST crystal for a 0.8 μm pump wavelength is indeed significantly lower than the $\chi_{111}^{OR} = 490$ pV/m, used for a 1.5 μm pump pulse. Moreover, at the crystal orientation corresponding to the $\chi_{111}^{OR}$ application the best phase-matching appears at the pump wavelengths between 0.70 and 0.72 μm, which are difficult to obtain directly from the Ti:sapphire amplifier.

To increase the THz output the use of a DAST/SiO$_2$ multilayer structure [Fig.1(b)] was recently proposed [18]. We should note, that this structure provides phase correction in a way which differs from the widely used quasi-phase-matching technique [19,20]. Alternating layers of DAST and quartz of appropriate thicknesses results in counteracting the phase mismatch that appears in individual layers, just because in DAST at the orientation providing the highest OR nonlinear susceptibility ($\chi_{111}^{OR}$) the group refractive index at 0.8 μm ($n_{0.8, gr}^{DAST} \approx 3.38$ [21]) is larger than the THz phase refractive index ($n_{THz, ph}^{DAST} \approx 2.4$ in the 2–8 THz range [22]), while in quartz $n_{0.8, gr}^{quartz} \approx 1.55$ for the ordinary rays $< n_{THz, ph}^{quartz} \approx 2.1$–2.2 (in the same THz range [23, 24]. The domain inversion in quasi-phase-matching structures [Fig.1(c)] results in a phase shift of 180 degrees. For this reason the proposed DAST/SiO$_2$ multilayer structure allows generation of near single THz pulses, whereas periodically poled quasi-phase-matching structures usually provide generation of multi-cycle THz pulses [20]. Due to the relatively small difference in refractive index between DAST and quartz, the reflection of the generated THz wave at the DAST/SiO$_2$ interface is about 0.2%. Whereas, the laser pump pulse reflection at the DAST/ SiO$_2$ interface is about 4.6%, which substantially limits the number of layers in the structure. The use of a DAST/SiO$_2$ multilayer structure allows employing the $\chi_{111}^{OR} = 1230$ pV/m of DAST [21] at 0.8 μm which is significantly higher than that at 1.5 μm (490 pV/m) [21]. In a previous paper [18], it was speculated that the use of femtosecond laser pulses at 800 nm rather than two monochromatic waves would provide generation of near-single-cycle THz pulses in a DAST/SiO$_2$ multilayer structure. However, the time profiles and spectra of THz pulses obtained with this method were not explicitly shown.

In this paper we present a profound numerical study on near-single-cycle THz pulse generation via collinear phase-corrected optical rectification of 800 nm femtosecond laser pulses in a series of different DAST/SiO$_2$ multilayer structures [Fig.1(b)], optimized for efficiency and for THz spectral output in different frequency ranges. The energy conversion efficiency as well as the temporal and spectral pulse shapes are presented for these structures [Fig.1(b)] and are compared with those produced in a bulk DAST crystal [Fig.1(a)] and in a quasi-phase-matching periodically poled DAST crystal [Fig.1(c)]. We show that modifications of the thicknesses of the DAST and SiO$_2$ layers allow broadband emission at variable bandwidth and tuning of the average frequency of the generated THz pulses in the frequency range from 3 to 6 THz while offering a pump-to-THz conversion efficiency about an order of magnitude higher than for bulk DAST pumped at 800 nm.

## 2. Model

The model developed for assessing the THz generation in the DAST/SiO$_2$ multilayer structure is based on the undepleted pump approximation. The energy losses of a laser pump pulse due to the reflection at the DAST/SiO$_2$ interfaces and the broadening of the laser pulse duration due to propagation in the DAST layers have been taken into account. At the orientation providing the highest OR nonlinear susceptibility the DAST crystal has relatively strong dispersion around 0.8 μm. According to our estimations, propagation of the 30 fs, 0.8 μm laser pulse though 90 μm of DAST results in increase of the laser pulse duration up to 75 fs. The laser pulse temporal broadening leads to a decrease in the generation efficiency and a shift of the THz pulse average frequency to the low frequency range. Recently this phenomenon was confirmed experimentally [17].

The nonlinear wave equation describing THz-wave generation by difference frequency mixing of the laser pump pulse Fourier components ($E(\omega_L + \omega_{THz})$ and $E(\omega_L)$):

$$\frac{\partial}{\partial z}E(z,\omega_{THz}) = -\frac{\alpha(\omega_{THz})}{2}E(z,\omega_{THz}) + i\frac{2\omega_{THz}d_{eff}}{n(\omega_{THz})c}\int_{\omega_{L2}}^{\omega_{L1}}E(\omega_L + \omega_{THz})E^*(\omega_L)\,e^{i\Delta k z}d\omega_L \quad (1)$$

was solved numerically. $E(z, \omega_{THz})$ is the Fourier component of THz pulse field at the frequency $\omega_{THz}$, $\alpha(\omega_{THz})$ and $n(\omega_{THz})$ are the THz absorption coefficient and refractive index of the DAST crystal, $d_{eff}$ is the coupling constant, and $\Delta k$ is the wave vector mismatch ($\Delta k = k(\omega_{THz} + \omega_L) - k(\omega_L) - k(\omega_{THz})$). All parameters of the DAST crystal ($\alpha(\omega_{THz})$, $d_{eff}$, $n(\omega_{THz})$, and $n(\omega_L)$) have been taken along the a-axis [21, 22]. The calculations were performed for individual Fourier components of THz pulse field in the frequency range of 0.7–10.5 THz. The final THz pulse shape was obtained by composition of the spectral components.

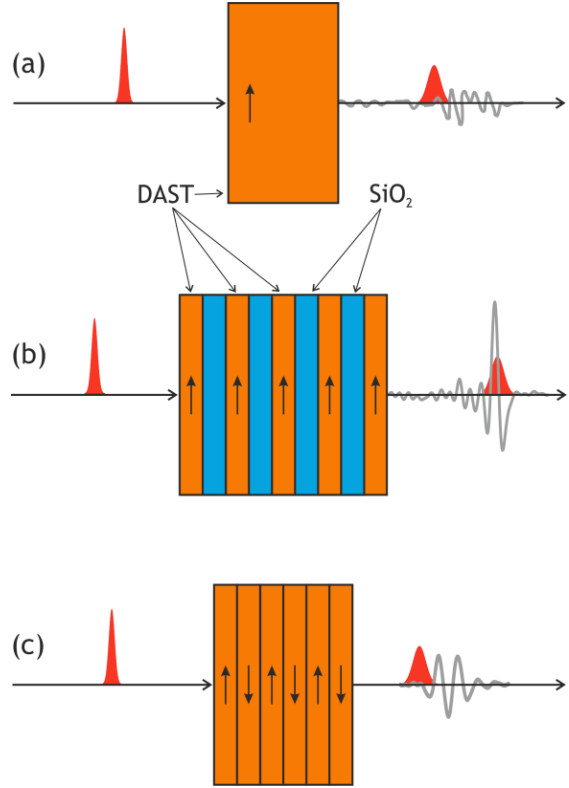

Fig. 1. THz generation via collinear optical rectification of 800 nm femtosecond laser pulses in studied samples: bulk DAST crystal (thickness ~ 100 μm) (a), multilayer DAST/SiO$_2$ structure (b), quasi-phase-matching periodically poled DAST crystal (c), see table 1 for details.

**Table.1 Parameters of the investigated multilayer DAST/SiO$_2$ structures (i-iii) and the quasi-phase-matching periodically poled DAST crystal (iv).**

| Str. | Layer number | Material | Thickness (μm) | Str. | Layer number | Material | Thickness (μm) |
|---|---|---|---|---|---|---|---|
| i | 1 | DAST | 13 | | 1 | DAST | 56.5 |
| | 2 | SiO$_2$ | 20 | | 2 | SiO$_2$ | 87 |
| | 3 | DAST | 12.5 | | 3 | DAST | 47.3 |
| | 4 | SiO$_2$ | 19 | iii | 4 | SiO$_2$ | 72 |
| | 5 | DAST | 12.5 | | 5 | DAST | 39.1 |
| | 6 | SiO$_2$ | 19 | | 6 | SiO$_2$ | 60.1 |
| | 7 | DAST | 15.6 | | 7 | DAST | 31.8 |
| ii | 1 | DAST | 20.5 | | 8 | SiO$_2$ | 49 |
| | 2 | SiO$_2$ | 32.5 | | 9 | DAST | 24 |
| | 3 | DAST | 21 | | 10 | SiO$_2$ | 37 |
| | 4 | SiO$_2$ | 32.5 | | 11 | DAST | 16.6 |
| | 5 | DAST | 20 | | 12 | SiO$_2$ | 25.6 |
| | 6 | SiO$_2$ | 32 | | 13 | DAST | 12.7 |
| | 7 | DAST | 19 | iv | 1 | DAST | 34 |
| | 8 | SiO$_2$ | 29.5 | | 2 | DAST | 34 |
| | 9 | DAST | 16.5 | | 3 | DAST | 34 |
| | 10 | SiO$_2$ | 25.5 | | | | |
| | 11 | DAST | 13.5 | | | | |
| | 12 | SiO$_2$ | 21 | | | | |
| | 13 | DAST | 10 | | | | |

For the quartz plates, we consider the orientation with the optical axis perpendicular to the plate face (ordinary ray geometry) because in this case its refractive index is not sensitive to the azimuth orientation of the quartz plates in the structure. This orientation of the quartz plates, to our mind, simplifies the DAST/SiO$_2$ multilayer structure manufacturing. As a downside, the ordinary ray absorption in the quartz crystal has a strong line at 7.9 THz, which is absent in the case of extraordinary ray absorption [24].

For the calculations, we used a 30 fs, 800 nm laser pulse, because it offers a pump pulse bandwidth large enough to support THz frequencies up to 10 THz by optical rectification. Today such laser pulses are delivered by commercially available high-power femtosecond laser systems. In order to maintain the validity of the undeleted pump approximation, a pump intensity of 1.5 GW/cm$^2$ was used, which is about 100 times less than the typical experimental value [12]. The calculations were performed for three different DAST/SiO$_2$ multilayer structures and for a DAST quasi-phase-matching sample. The properties of the studied multilayer structures are summarized in Table 1.

The thickness of a DAST layer is optimized in order to get desired THz signal output (either maximal intensity, either central frequency). In order to get maximal conversion efficiency regardless of frequency [case ii in Fig.4] the thickness of each DAST layer is determined as the position where phase mismatch starts to be dominating and THz intensity starts decreasing. On the other hand, when a THz wave with a particular central frequency is sought [cases i and iii in Fig.4], the thickness of each DAST layer is determined iteratively as the optimal tradeoff between conversion and position of the output spectrum. Different thickness for the SiO$_2$ layers were initially tested, corresponding to SiO$_2$/DAST values ranging from 1.4 to 1.8. The value selected (1.54) corresponds to the best phase matching between fundamental wave and the THz wave at the beginning of the next DAST layer in the frequency range of interest. The thicknesses of the DAST and SiO$_2$ layers in a structure are changing because of the decrease of the laser pulse intensity due to the reflection at the DAST/SiO$_2$ interfaces and the temporal pulse broadening.

## 3. Results and discussion

The time profile and spectrum of the THz pulse generated by multilayer structures (i-iii) and by a quasi-phase-matching periodically poled DAST crystal (iv) are shown in Fig. 2 and Fig. 3, respectively. The vertical lines in Fig. 3 show the average frequencies of the spectra.

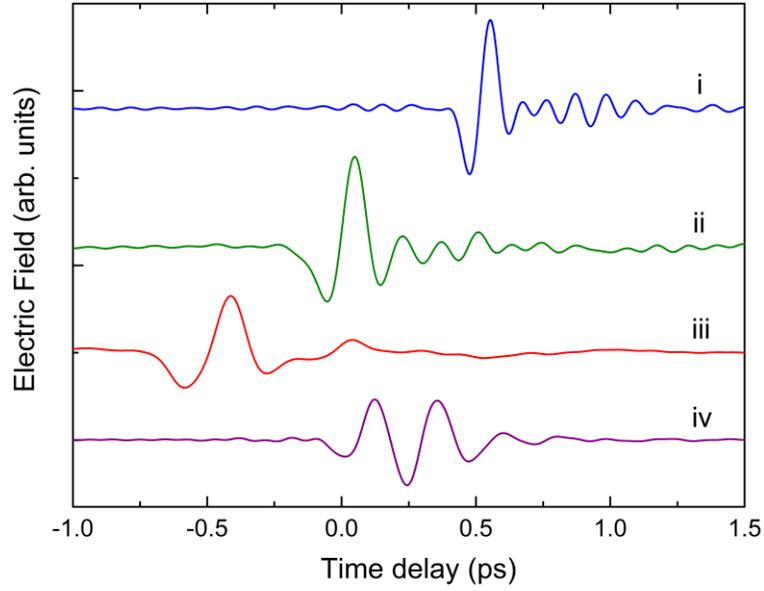

Fig. 2. Temporal profiles of THz pulses calculated for three different DAST/SiO$_2$ multilayer structures (curves i, ii, and iii; see text for details) and for a quasi-phase-matching periodically poled DAST crystal (curve iv).

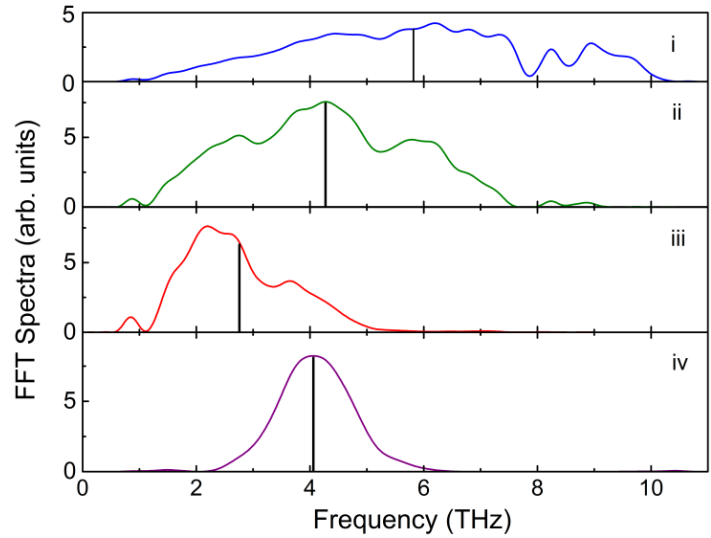

Fig. 3. FFT spectra of THz pulses calculated for three different DAST/SiO$_2$ multilayer structures (i, ii, and iii; see text for details) and for a quasi-phase-matching periodically poled DAST crystal (iv).

Figure 4 represents the laser-to-THz energy conversion efficiency as a function of laser pulse propagation distance, calculated for three different DAST/SiO$_2$ multilayer structures (traces i, ii, and iii), a quasi-phase-matching periodically poled DAST crystal (trace iv), and a bulk DAST crystal (trace v). In the bulk DAST crystal, the laser-to-THz energy conversion efficiency begins to decrease after a propagation distance of 15 μm, because of a lack of phase-matching. In order to overcome this decrease of the laser-to-THz energy conversion efficiency, we suggest the use of a DAST/SiO$_2$ multilayer structure, which is capable of counteracting the phase mismatch appearing in different layers. The step-like periodic decrease of the energy conversion efficiency shown in Fig. 4 (curves i and ii) occurs due to absorption of the generated THz waves in the quartz layers.

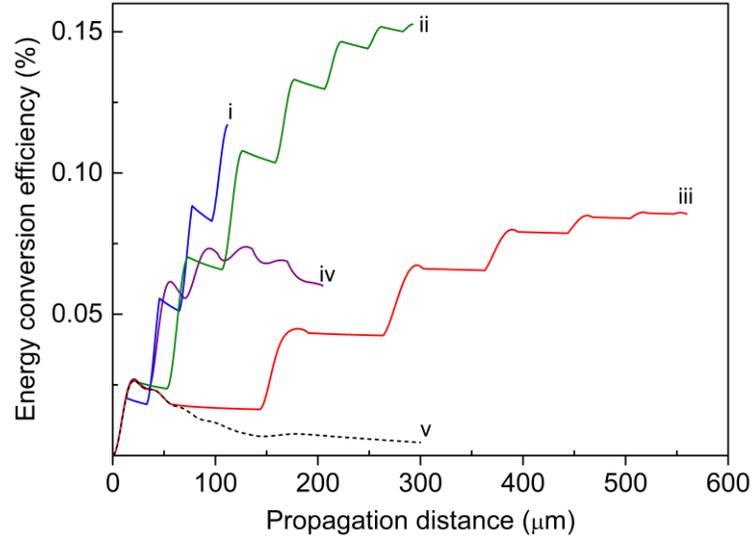

Fig. 4. Laser-to-THz energy conversion efficiency as a function of laser pulse propagation distance calculated for three different DAST/SiO$_2$ multilayer structures (traces i, ii, and iii), a quasi-phase-matching periodically poled DAST crystal (trace iv), and a bulk DAST crystal (trace v) pumped with the output of a Ti:sapphire laser (0.8 μm).

The first DAST/SiO$_2$ structure (i) was optimized for obtaining near-single-cycle pulses with an average frequency of 6 THz. In this case we obtained the broadest THz spectrum covering the entire range of 1-10 THz [curve i in Fig. 3]. The strong dips in the pulse spectrum at 7.9 and 8.5 THz appear due to absorption lines in the quartz and DAST crystals, respectively. The resulting temporal shape is near to single-cycle [curve i in Fig. 2]. Due to the small number of DAST layers the conversion efficiency is limited but still 4.5 times higher than that of a 15 μm bulk DAST crystal [Fig. 4, curves i and v]. A larger conversion efficiency can be achieved by stacking a larger number of DAST/SiO$_2$ layers (structure ii). However, an increase in the number of structure layers results in a shift of the average THz frequency lower than 6 THz owing to the temporal broadening of the femtosecond laser pulses propagating in the DAST layers.

The second DAST/SiO$_2$ multilayer structure (ii) was optimized for obtaining the highest laser-to-THz energy conversion efficiency. In this case, we obtain near-single-cycle THz pulses [curve ii in Fig. 2] with an average frequency of 4.3 THz [curve ii in Fig. 3]. This structure provides conversion efficiency about 6 times higher than a 15 μm bulk DAST crystal [curves ii and v in Fig. 4].

The third DAST/SiO$_2$ multilayer structure (iii) was designed for the generation of near-single-cycle THz pulses with an average frequency of about 3 THz [curve iii in Fig.3]. The structure provides the conversion efficiency about 2.5 higher than that obtained with a 15 μm bulk DAST crystal [curves iii and v in Fig. 5].

Finally, the time profile and spectrum of the THz wave generated in a quasi-phase-matching DAST crystal consisting of three periodically polled layers are show in Fig. 2 (curve iv) and Fig. 3 (curve iv), respectively. The thickness of the polled DAST layers was chosen to obtain efficient generation at 4 THz.

Presently, manufacturing a DAST/SiO$_2$ multilayer structure is a difficult task, mainly because of the problems related with growing large-aperture thin (10 − 20 μm) DAST crystal layers with well controlled thicknesses and crystallographic orientation. However, due to the increasing demand, the technology of nonlinear organic crystal growth is developing fast. Growing of 3 μm single crystal films of DAST with an area of 30–40 mm$^2$ on a quartz substrate using a modification of the shear method was reported [25]. From the paper it remains unclear whether this method allows growing thinker DAST films or not. One of the methods for manufacturing the DAST/SiO$_2$ multilayer structure can be growing a few DAST/SiO$_2$ double layers with required thicknesses and then stacking them together in vacuum in order to avoid air bubbles formation at the DAST/SiO$_2$ interfaces. Recently the growth of 23 μm film of organic salt crystal DASC and its application for THz generation was demonstrated [26]. DASC has optical properties similar to DAST and could be used for multilayer structure manufacturing as well.

## 4. Conclusion

We have presented a model calculation showing that DAST/SiO$_2$ multilayer structures can be used for the efficient generation of near-single-cycle electromagnetic pulses in the 2–9 THz frequency range via collinear optical rectification of 800 nm femtosecond laser pulses. According to our calculations, the proposed structure provides higher laser-to-THz energy conversion efficiency than a bulk and quasi-phase-matching periodically poled DAST crystals pumped at Ti:sapphire wavelength. Tailoring the multi-layer structure gives novel opportunities to spectral shaping the THz radiation. Once the technological hurdles in productions are overcome, we are convinced that the proposed scheme offers a simple way for obtaining high energy near-single-cycle THz pulses with average frequency tunable in a frequency range which is difficult of access today.

## Acknowledgments

This work was supported by NCCR Molecular Ultrafast Science and Technology (NCCR MUST), a research instrument of the Swiss National Science Foundation (SNSF). A. G. S. acknowledges the support of the Marie Curie Action for International Incoming Fellowships (FP7-PEOPLE-2010-IIF), under the HINTS project, C. P. H. by SNF (Grant No. PP00P2_128493).